\documentclass[aps,pra,twocolumn,tightenlines,superscriptaddress,amsmath,amssymb, 10pt]{revtex4-1}
\pdfoutput=1
\usepackage{graphicx}
\usepackage{bm}
\usepackage{accents}
\usepackage{multirow}
\usepackage{grffile}
\usepackage{amsfonts}
\usepackage{mathtools}

\usepackage{xcolor}
\definecolor{dark-red}{rgb}{0.4,0.15,0.15}
\definecolor{dark-blue}{rgb}{0.15,0.15,0.4}
\definecolor{medium-blue}{rgb}{0,0,0.5}
\usepackage{hyperref}
\usepackage[all]{hypcap}%fix problems with figures and tables of the 'hyperref' package
\usepackage{cleveref}% Package hperref works wrongly within the 'subequations' environment without this.
\hypersetup{
    colorlinks, linkcolor={dark-red},
    citecolor={dark-blue}, urlcolor={medium-blue}
} %These setups replace the ugly and distracting red and green boxes of the 'hyperref' package with colored texts.
 \newcommand{\be}{\begin{equation}}
 \newcommand{\ee}{\end{equation}}
\newcommand{\bea}{\begin{eqnarray}}
\newcommand{\eea}{\end{eqnarray}}
\newcommand{\ba}{\begin{eqnarray*}}
\newcommand{\ea}{\end{eqnarray*}}

\usepackage{multirow}

\newcommand{\half}{\frac{1}{2}}
\newcommand{\ket}[1]{\lvert\, #1\, \rangle}
\newcommand{\bra}[1]{\langle\, #1\, \rvert}
\newcommand{\braket}[2]{\langle\, #1\,\vert\, #2 \,\rangle}

\newcommand{\norm}[1]{\lvert #1 \rvert}
\DeclareMathOperator{\tr}{Tr}

\newcommand{\erg}{\varepsilon}

\newcommand{\g}{g}

\newcommand{\mm}{{m}}

\newcommand{\action}{{\mathcal A}}
\newcommand{\sB}{\mathcal{B}}
\newcommand{\dif}{d}
\newcommand{\ssp}{\hspace{0.4pt}}
\newcommand{\normb}[1]{\big\lvert #1 \big\rvert}
\newcommand{\proj}[1]{\ket{#1}\!\bra{#1}}
\newcommand{\dt}[1]{\accentset{\vspace{0.5pt}\hspace{0.5pt}\mbox{\large .}}{#1}} %

\newcommand{\nrepl}{R}
\newcommand{\trprop}{\mathcal V}

\newcommand{\OB}{\mathrm{\scalebox{0.6}{OB}}}
\newcommand{\MF}{\mathrm{\scalebox{0.6}{MF}}}
\newcommand{\QMC}{\mathrm{\scalebox{0.6}{QMC}}}
\newcommand{\ellopen}{\ell}
\newcommand{\mfield}{\lambda}

\newcommand{\ratio}{\varrho}
\newcommand{\tauprim}{\tau_1}
\newcommand{\entropy}{Q}
\newcommand{\free}{{\mathcal F}}
\newcommand{\xiplus}{\zeta}
\newcommand{\eat}[1]{}

\def\ket#1{| #1 \rangle}
\def\bra#1{\langle #1 |}
\def\braket#1#2{\langle #1 | #2 \rangle}

\def\B{{\rm B}}

\def\({\left(}
\def\){\right)}

\usepackage{color}
\usepackage[english]{babel}
\usepackage[caption=false]{subfig}

\begin{document}

\title{\bf Path-Integral Quantum Monte Carlo Simulation with Open-Boundary Conditions}
\date{\today}
\author{Zhang Jiang}
\affiliation{QuAIL, NASA Ames Research Center, Moffett Field, California 94035, USA}
\affiliation{SGT Inc., 7701 Greenbelt Rd., Suite 400, Greenbelt, MD 20770}
\author{Vadim N. Smelyanskiy}
\affiliation{Google, Venice, CA 90291, USA}
\author{Sergio Boixo}
\affiliation{Google, Venice, CA 90291, USA}
\author{Hartmut Neven}
\affiliation{Google, Venice, CA 90291, USA}
\begin{abstract}
The tunneling decay event of a metastable state in a fully connected quantum spin model can be simulated efficiently by path-integral quantum Monte Carlo (QMC) [Isakov \textit{et al.}, \href{https://doi.org/10.1103/PhysRevLett.117.180402}{Phys.\ Rev.\ Lett.\ \textbf{117}, 180402 (2016)}.]. This is because the exponential scaling with the number of spins of the thermally assisted quantum tunneling rate and the Kramers escape rate of QMC are identical [Jiang \textit{et al.}, \href{https://doi.org/10.1103/PhysRevA.95.012322}{Phys.\ Rev.\ A \textbf{95}, 012322 (2017)}.], a result of a dominant instantonic tunneling path. In Isakov \textit{et al.}, it was also conjectured that the  escape rate in open-boundary QMC is quadratically larger than that of conventional periodic-boundary QMC; therefore, open-boundary QMC might be used as a powerful tool to solve combinatorial optimization problems. The intuition behind this conjecture is that the action of the instanton in open-boundary QMC is a half of that in periodic-boundary QMC. Here, we show that this simple intuition---although very useful in interpreting some numerical results---deviates from the actual situation in several ways. Using a fully connected quantum spin model, we derive a set of conditions on the positions and momenta of the end points of the instanton, which remove the extra degrees of freedom due to open boundaries. In comparison, the half-instanton conjecture incorrectly sets the momenta at the end points to zero. We also found that the instantons in open-boundary QMC correspond to quantum tunneling events in the symmetric subspace (maximum total angular momentum) at all temperatures, whereas the instantons in periodic-boundary QMC typically lie in subspaces with lower total angular momenta at finite temperatures. This leads to a lesser-than-quadratic speedup  at finite temperatures. The results provide useful insights in utilizing open-boundary QMC to solve hard optimization problems. We also outline the generalization of the instantonic tunneling method to many-qubit systems without permutation symmetry using spin-coherent-state path integrals.
\end{abstract}

\maketitle

\section{Introduction}

Computational hard combinatorial optimization problems can be
mapped to spin glass models in statistical
physics~\cite{fu_application_1986}. The energy
landscapes of the corresponding problem Hamiltonians $H_P$ are rough and posses large numbers of spurious local minima.  % Their size grows with the problem size $N$ (number of spins).
Conventional optimization strategies, such as simulated annealing, exploit thermal over--the--barrier transitions to escape from the local minima. In quantum annealing (QA)~\cite{PhysRevE.58.5355,*brooke_quantum_1999,Farhi20042001,Santoro29032002,boixo2014, ronnow_defining_2014, King:2015vl}, tunneling offers additional paths for systems to go to low-energy states~\cite{boixo_computational_2014}. In an archetypical example of QA, the state evolution is determined by a time-dependent Hamiltonian $H(t) = H_P - \Gamma(t) \sum_j \sigma_j^x$, where $\sigma_j^x$ is the Pauli-$x$ operator of the $j$th spin, and $\Gamma(t)$ slowly interpolates from a large value to $0$. All low-energy eigenstates of $H(t)$ are localized in the vicinity of
the minima of $H_P$ at sufficiently small values of $\Gamma$~\cite{Altshuler13072010}. The order of the energies of two minima changes at an avoided crossing, and quantum tunneling between the two minima gives rise to the energy gap $\Delta$. The quantum state remains pure in the absence of an environment, and it closely follows the instantaneous ground state of $H(t)$ when $\Gamma(t)$ is changed slowly enough. Such a process is called adiabatic quantum computation~\cite{Farhi20042001}, where the state dynamics can be described as a cascade of Landau--Zener transitions at the
avoided crossings \cite{Santoro29032002}.
 
Interaction with an environment causes relaxation, and this can suppress tunneling. On the other hand, the environment also gives rise to thermal excitations to higher energy levels from where the system can tunnel faster~\cite{kechedzhi_open-system_2016}. This is called thermally assisted
tunneling~\cite{larkin1983pis,*garanin1997thermally,*affleck_quantum-statistical_1981}, and it was recently discussed in QA with flux qubits~\cite{amin_macroscopic_2008,dickson_thermally_2013}. We assume that the system has a free-energy minimum associated with a thermodynamically metastable state and that the incoherent tunneling decay rate $W$ of this state is much smaller than the smallest  relaxation rate  towards the quasiequilibrium distribution in the domain associated with this state. Then the  incoherent tunneling decay rate is $W \propto  \Delta^2/(\hbar^2\gamma)$, where $\gamma \gg \Delta/\hbar$ is a largest relaxation rate (typically,  dephasing rate).

Path-integral quantum Monte Carl (QMC) is a classical Markov chain Monte Carlo algorithm to calculate thermodynamic properties of quantum systems; it is the most efficient algorithm to compute exponentially small gaps at first order phase transitions~\cite{young2010first,hen2011exponential,boixo2014} and is often used to simulate QA in spin glasses~\cite{Santoro29032002,PhysRevB.66.094203,*martonak_quantum_2004,*battaglia_optimization_2005,*santoro_optimization_2006,boixo2014,Heim12032015}. 
The equivalence between the exponential scaling of the QMC transition rate and the QA tunneling rate in a fully connected quantum spin model was established numerically at the effective ``zero-temperature" limit~\cite{isakov_understanding_2016}, where the effect of thermal excitations on the tunneling rate can be neglected. A detailed theoretical calculation on this equivalence based on an instanton technique was given in~\cite{jiang_scaling_2017} at finite temperatures. Therein, the instanton method was used to demonstrate the identical exponential scaling of the QA tunelling rate and the QMC escape rate for a bit-symmetric cost function with a thin, high energy barrier (\lq\lq Hamming weight with a spike''). Crosson and Harrow also considered the same cost function, and they derived a bound on the mixing time of the underlying Markov chain~\cite{crosson_simulated_2016}; their results also showed that QMC takes polynomial time to find the solution in the regime with no tunneling.

In QMC, paths are being sampled instead of the actual configurations of the system. A path consists of
a sequence of replicas of the system, where changes in configurations between adjacent replicas are penalized energetically. The Kramers escape event in the stochastic process describes the transition from a local minimum to the global minimum, which is dominated by a single \lq\lq transition state" (a saddle point) that the system needs to reach in order to make an escape from the metastable state. As we show (see also Refs.~\cite{isakov_understanding_2016, jiang_scaling_2017}), this transition state corresponds to an instanton, and the change in free energy needed to reach this state is the same as that in the quantum tunneling case. This explains the equivalence in the exponential scaling of the QMC transition rate and thermally assisted quantum tunneling rate. 

In Ref.~\cite{isakov_understanding_2016}, the authors found numerically that a quadratic speedup may be achieved in QMC escape events by using open-boundary condition (OBC) instead of periodic-boundary condition (PBC); it was conjectured that the speedup is due to an escape path of a half instanton. Physical quantities usually cannot be predicted correctly by using open-boundary QMC; however, it can be used as a physics-inspired classical algorithm for combinatorial optimization problems. In a recent numerical study~\cite{mazzola_accelerated_2016}, it was found that the autocorrelation time in QMC simulations with OBC is significantly smaller than that of conventional QMC methods. 

Because QMC samples paths instead of the physical states, it might have conserved quantities not present in the physical system, such as the number of world lines (particles, magnetization), braiding, or winding numbers~\cite{evertz_loop_2003,hastings}. Topological obstructions can arise from these conserved quantities, which prevent QMC from reaching equilibrium even though the quantum Hamiltonian is gapped. Open-boundary QMC is not immune to this, but it can escape many topological obstructions that trap periodic-boundary QMC. Recently, Andriyash and Amin~\cite{andriyash_can_2017} argued that topological obstructions can make QMC less efficient than QA when there are multiple tunneling paths between two minima. These obstructions partially prevent the forming of instantons in periodic-boundary QMC by creating a barrier in the middle. However, open-boundary QMC is not encumbered by such obstructions.

Here, we consider open-boundary QMC of a fully connected spin model with a bit-symmetric cost function. We show that the saddle point of the mean-field QMC free-energy functional satisfies the same differential equations as those in periodic-boundary QMC. We also derived a set of conditions on the positions and momenta of the end points of instantons in open-boundary QMC, which uniquely determine the instanton solution by removing the extra degrees of freedom due to OBC. Interestingly, we found nonzero initial and final momenta of instanton in open-boundary QMC. This can be attributed to the extra entropic factors at the boundaries. The free-energy at the local minimum also corresponds to a nonstationary solution in the path-integral formalism due to the same entropic factors. Another somewhat surprising fact about open-boundary QMC is that the optimal tunneling path lies in the symmetric subspace (maximum total angular momentum) at all temperatures. In contrast, the optimal tunneling paths in periodic-boundary QMC typically have total angular momentum less than the maximum value at finite temperatures. Our analytical results show that the conjecture in Ref.~\cite{isakov_understanding_2016} is only approximately correct: The QMC escape rate is not always enhanced quadratically by using open-boundary conditions.
  
In Sec.~\ref{sec:open_boundary} we formally establish the free-energy functional for QMC with OBC and derive its saddle-point equation. In Sec.~\ref{sec:free_erg} we 
compute the free-energy functional of the open-boundary QMC at the saddle point using the instanton approach. In Sec.~\ref{sec:wkb} we map the open-boundary QMC free-energy functional to a quantum propagator and use the Wentzel-Kramers-Brillouin (WKB) approach to derive the corresponding ``tunneling rate'' in the model. The result conforms with the QMC escape rate derived in preceding sections.

\section{Saddle points of path integral QMC with OBC}
\label{sec:open_boundary}

We consider a fully connected quantum system of $N$ spins~\cite{Semerjian-wkb,kechedzhi_open-system_2016},
\begin{align}
 \hat H = -2 \Gamma S_x - N g(2 S_z/N), \quad S_\alpha
  = \frac 1 2 \sum_{j=1}^{N} \sigma_{\alpha}^{(j)}  \label{eq:H-sigma}\,,
\end{align}
where $\Gamma$ is the strength of the transverse field, $\sigma_{\alpha}^{(j)}$ is the Pauli matrix of the $j$th spin with $\alpha=x,y,z$, $S_\alpha$ is the $\alpha$ component of the total spin operator, and $g$ is an arbitrary function.

A classical method to calculate the partition function $ Z = \tr e^{-\beta \hat H}$
is the path-integral QMC, where the \lq\lq imaginary-time'' interval $[0,\,\beta]$ is sliced into $\nrepl$ pieces. In the limit $\nrepl\rightarrow \infty$, we have
\begin{align}\label{eq:partition_f_pimc}
\begin{split}
 Z = &\hspace{-10pt}\sum_{\underline\sigma(0),\ldots,\underline\sigma(\nrepl-1)}\bra{\underline\sigma(0)}  e^{-\Delta \hat H}\proj{\underline\sigma(1)} e^{-\Delta \hat H} \\
 &\quad\cdots\proj{\underline\sigma(\nrepl-1)} e^{-\Delta \hat H}\ket{\underline\sigma(0)}\,,
\end{split}
 \end{align}
where $\Delta =\beta/\nrepl$ and $\underline\sigma(\tau) = \{\sigma_j(\tau)\}_{j=1}^N$. 
Equation~(\ref{eq:partition_f_pimc}) also takes the form
\begin{align}\label{eq:partition_func}
  Z = \tr e^{-\beta H_{\scriptscriptstyle\QMC}}\,,
\end{align}
where $H_\QMC$ is a Hamiltonian on $\nrepl \times N$ classical spins,
\begin{align}\label{eq:hamiltonian}
 H_\QMC =-J\sum_{j=1}^{N}\sum_{\tau=0}^{\nrepl-1}\sigma_j(\tau)\sigma_j(\tau+1) -\frac{N}{\nrepl}\sum_{\tau = 0}^{\nrepl-1} \g[\mm(\tau)]\,.
\end{align}
%The convention $\underline\sigma(\nrepl-1)= \underline\sigma(0)$ is often used, which is a consequence of the trace in the partition function. 
The quantity $\mm(\tau)$ in Eq.~\eqref{eq:hamiltonian} is the total magnetization,
\begin{align}\label{eq:m_tau}
 \mm(\tau)= \frac{1}{N}\sum_{j=1}^{N} \sigma_j(\tau)\,,
\end{align}
and $J$ is the effective coupling strength between adjacent replicas,
\begin{align}
J=-\frac{1}{2\beta} \ln \tanh\big(\Gamma\Delta\big)\geq 0\,.
\end{align}
When the coupling strength $J=\infty$ ($\Gamma\rightarrow 0$), we have $\mm(\tau)=\mm$ for $\tau =0,\ldots, \nrepl-1$; the free-energy functional $\free$ then reduces to the classical case,
\begin{align}
 \free_\mathrm{classical}(\mm) = - \g (\mm)-\frac{\entropy(\mm)}{\beta} \,,
\end{align}
where $\entropy$ is the binary entropy,
\begin{align}\label{eq:entropy}
 \entropy(\mm) &\equiv -\frac{1+\mm}{2}\ln\frac{1+\mm}{2} -\frac{1-\mm}{2}\ln\frac{1-\mm}{2}\;.
 %\\ &= -\frac{1}{2}\ln(1-m^2)-\frac{m}{2}\ln\frac{1+m}{1-m} +\ln 2\,,
\end{align}
% When $J=0$ ($\Gamma\rightarrow \infty$), the free-energy functional of a trajectory $\mm(0),\ldots,\mm(\nrepl-1)$ takes the form
% \begin{align}
% \m\free =-\frac{\ln Z}{\beta N}= -\sum_{\tau = 0}^{\nrepl-1} \bigg(\frac{\g[\mm(\tau)]}{\nrepl} +\frac{\entropy[\mm(\tau)]}{\beta} \bigg)\,,
% \end{align}

In Appendix~\ref{sec:PIQMC}, we review the mean-field approach to path-integral QMC with periodic and open-boundary conditions. Here, we discuss how to calculate the saddle point of the free-energy functional of the QMC Hamiltonian~(\ref{eq:hamiltonian}) with open-boundary condition. The free-energy functional of open-boundary QMC takes the following form in the continuous limit $\nrepl\rightarrow \infty$ (see Appendix~\ref{sec:PIQMC}):
\begin{align}\label{eq:F_OB}
\begin{split}
\free_\OB=\frac{1}{\beta} \int_{0}^{\beta}\Big(\mm(\tau) \lambda(\tau)-g[\mm(\tau)]\Big) d\tau
- \frac 1 \beta \ln \widetilde\trprop(\lambda)\,,
\end{split}
\end{align}
where $\lambda(\tau)$ is a to-be-determined function of the imaginary time $\tau\in[0,\,\beta]$, and
\begin{align}\label{eq:ob_partition_f}
  \widetilde\trprop(\lambda) =\tr \big(\omega K^{\beta,\, 0}( \lambda) \,\big)\,,\quad \omega=1+\sigma_x=\begin{pmatrix}1&1\\1&1\end{pmatrix}\,.
\end{align}
The propagator $K^{\beta,0}$ corresponds to a spin-1/2 particle evolving in imaginary time under the action of the time-dependent magnetic field $\mathbf B(\tau)$,
\begin{gather}
% \widetilde\trprop(\bm \lambda) =\tr \big(\omega K^{\beta,0}\big),\label{eq:Lambda}\\
K^{\tau_2,\tau_1} ={\rm T_{+}}e^{-\int_{\tau_1}^{\tau_2} d\tau  H_0(\tau)}\,,\label{eq:propagator}\\
H_0(\tau) =-\mathbf{B}(\tau)\cdot \bm{\sigma},\quad \mathbf B(\tau)=\big(\Gamma,0,\lambda(\tau)\big)\label{eq:H0}\;,
\end{gather}
where ${\rm T_{+}}$ is the time-ordering operator, and $\bm \sigma=(\sigma_x,\sigma_y,\sigma_z)$ is the vector of Pauli matrices.

The saddle-point conditions for the free-energy functional~(\ref{eq:F_OB}) are
\begin{equation}\label{eq:dm-var}
\lambda(\tau)= g^\prime[\mm(\tau)]\,,\quad \mm(\tau)=\frac{\delta \ln \widetilde\trprop(\lambda)}{\delta \lambda(\tau)}\,.
\end{equation}
To solve these equations, we introduce a vector function ${\bf  m}(\tau)=\big(m_x(\tau),m_y(\tau),m_z(\tau)\big)$ 
corresponding to the following expectation values of the Pauli operators,
\begin{align}\label{eq:dm-vec}
\begin{split}
  {\bf  m}(\tau) &= \frac{{\rm Tr} \left (\omega K^{\beta,\tau}  {\bm \sigma}
  K^{\tau,0}\right )}{{\rm Tr}\,\big(\omega K^{\beta,0}\big) } \\
  &= \frac{\delta }{\delta {\bf B}(\tau)} \ln {\rm Tr}\big(\omega
  K^{\beta,0}\big)\;,
  \end{split}
\end{align}
where $K^{\tau_2,\tau_1}$ is defined in  Eq.~(\ref{eq:propagator}) and ${\bf B} (\tau)$ in Eq.~\eqref{eq:H0}. Differentiating Eq.~\eqref{eq:dm-vec} with respect to $\tau$ and using (\ref{eq:H0}), we obtain
\begin{equation}
\frac{d{\bf m}}{d\tau} =\frac{{\rm Tr} \left (\omega K^{\beta,\tau} [H_0(\tau), \bm{\sigma}] K^{\tau,0}\right )}{{\rm Tr}\,\big(\omega K^{\beta,0}\big) }\;,\label{eq:mvec}
\end{equation}
which can be rewritten into the form
\begin{equation}
\frac{d{\bf m}}{d\tau}=2i {\bf B}\times {\bf m} = -2 i \frac{\partial {\mathcal H_0}({\bf
    m})}{\partial {\bf m}} \times {\bf m}\,, \label{eq:dmvec}
\end{equation}
where
\begin{equation}
 {\mathcal H_0}({\bf m}) = -\Gamma m_x-g(m_z)\label{eq:H0c}\;.
 \end{equation}
In component form, Eq.~\eqref{eq:dmvec} becomes
\begin{gather}
\dt m_x = -2\lambda i m_y\,,\label{eq:dt_mx}\\
i\dt m_y = -2\lambda m_x + 2\Gamma m_z\,,\label{eq:dt_my}\\
\dt m_z =  2 \Gamma i m_y .\label{eq:dt_mz}
\end{gather}
Equation~(\ref{eq:dmvec}) allows for two integrals of motion,
\begin{gather}
{\cal H}_{0}({\bf m})=\erg,\label{eq:int1}\\
{\bf m}(\tau)\cdot {\bf m}(\tau)=\ellopen^2 \label{eq:int2}
\end{gather}
where  ${\bf m}(\tau)\cdot {\bf m}(\tau)\equiv m_x^2(\tau)+m_y^2(\tau)+m_z^2(\tau)$.
Using these two integrals of motion, we have
\begin{gather}
m_x =-\frac{\erg+g(m_z)}{\Gamma} =\sqrt{\ellopen^2-m_z^2}\cosh p\;,\label{eq:mx}\\
i m_y = \sqrt{\ellopen^2-m_z^2} \sinh  p\;,\label{eq:my}
\end{gather}
where $p$ serves as the \lq\lq conjugate momentum'' to $\mm_z$. Solving $m_y$ as a function of $m_z$ from Eqs.~\eqref{eq:mx} and \eqref{eq:my} and putting the result into Eq.~\eqref{eq:dt_my}, we have the instanton equation
\begin{align}\label{eq:instanton_eq}
 \norm{\dt m_z} &=  2 \sqrt{[\erg+g(m_z)]^2-\Gamma^2 (\ellopen^2-m_z^2)}\,.
\end{align}
We still need to specify $\erg$, $\ellopen$, and the initial condition of $\mm_z$ to fully determine the instanton trajectory based on the differential equation~\eqref{eq:instanton_eq}.  Using the identity $\sigma_x\omega = \omega\sigma_x =\omega$, we have
\begin{align}\label{eq:_open_initial_condition}
 \tr (\omega K^\beta \sigma_x) = \tr (\omega K^\beta)\,,\quad  i\tr (\omega K^\beta \sigma_y) = \tr (\omega K^\beta \sigma_z)\,.
\end{align}
As a result, we have the following initial conditions:
\begin{align}\label{eq:initial_instaton}
m_x(0) = 1\,,\quad  m_z(0) = i m_y(0)  = \frac{1}{2\Gamma}\,\dt m_z(0)\,.
\end{align}
Because $\ellopen$ is a constant of motion, its value can be determined at $\tau=0$,
\begin{align}
\ellopen^2 = \sum_{j=x,y,z}m^2_j(0) = 1\,.
\end{align}
This means that the parameter $\ellopen$ is independent of $\beta$ under OBC; the same condition can also be derived using the WKB approach [see Eq.~\eqref{eq:wkb_partition_f}].  Similarly, we have
\begin{align}\label{eq:end_instanton}
m_x(\beta) = 1\,,\quad  m_z(\beta) = -i m_y(\beta) = -\frac{1}{2\Gamma}\,\dt m_z(\beta)\,.
\end{align}
From the energy conservation condition Eq.~(\ref{eq:int1}) and the condition $m_x(0) =m_x(\beta) = 1$, we have
\begin{align}
g[m_z(0)] = g[m_z(\beta)]\,.
\end{align}
This condition relates the position of the end points of the instanton trajectory under OBC. 

\section{Free energy of the instanton}
\label{sec:free_erg}

To calculate the free energy from Eq.~\eqref{eq:F_OB}, it remains to evaluate the quantity
\begin{align}\label{eq:trace_OB}
 \widetilde\trprop = \tr( \omega K^{\beta, 0}) = 2\,\bra{+}K^{\beta, 0}\ket{+} \,,
\end{align}
where $\ket{+} = \frac{1}{\sqrt 2}(\ket{0}+\ket{1})$. We introduce the double propagator $K^{\tau,0}\otimes K^{\tau,0}$ acting on the original qubit and a replica qubit, which is generated by the Hamiltonian
\begin{align}
H_0^{(2)} = H_0\otimes I + I\otimes H_0\,,
\end{align}
where $H_0 =-\Gamma \sigma_x-\g'(\mm_z) \sigma_z$ is defined in Eq.~(\ref{eq:H0}).  With the Bell basis $\ket{\Phi^+} = \frac{1}{\sqrt{2}}(\ket{00}+ \ket{11})\,,\;
\ket{\Phi^-} = \frac{1}{\sqrt{2}} (\ket{00}- \ket{11})\,,\;
\ket{\Psi^+} = \frac{1}{\sqrt{2}} (\ket{01}+ \ket{10})\,,\;
\ket{\Psi^-} = \frac{1}{\sqrt{2}} (\ket{01}- \ket{10})$, we have
\begin{align}
&-H_0^{(2)} \ket{\Phi^+} = 2 \Gamma \ket{\Psi^+} + 2 \g'(\mm_z) \ket{\Phi^-}\;,\label{eq:Phi+}\\
&-H_0^{(2)}\ket{\Phi^-} = 2 \g'(\mm_z) \ket{\Phi^+}\;,\label{eq:Phi-}\\
&-H_0^{(2)}\ket{\Psi^+} = 2 \Gamma \ket{\Phi^+}\;,\label{eq:Psi+}\\
&-H_0^{(2)}\ket{\Psi^-} = 0\;.
\end{align}
Because the time evolution is closed in the symmetric subspace of the two qubits, we have
\begin{align}\label{eq:Xi}
\begin{split}
 \ket{\bm\xi(\tau)} &= K^{\tau, 0}\otimes K^{\tau, 0}\ket{\bm\xi(0)}\\
 &=-\xi_x (\tau)\ket{\Phi^-} +i \xi_y (\tau)\ket{\Phi^+} +  \xi_z(\tau)\ket{\Psi^+}\;,
 \end{split}
\end{align}
where $\xi_x, i\xi_y$, and $\xi_z$ take real values.
%\zj{The sign of the term $i \xi_y (\tau)\ket{\Phi^+}$ in the above equation has been changed.}
According to Eqs.~(\ref{eq:Phi+})--(\ref{eq:Psi+}), the state vector $\boldsymbol \xi = (\xi_x,\xi_y,\xi_z)$ satisfies the linear differential equation
\begin{equation}
\frac{d{\boldsymbol \xi}}{d\tau}=-2 i \frac{\partial {\mathcal H_0}({\bf
    m})}{\partial {\bf m}} \times {\boldsymbol \xi} \;,\label{eq:dxivec}
\end{equation}
where $\bf m$ is determined by the instanton solution Eqs.~(\ref{eq:mx})--(\ref{eq:instanton_eq}).
%  Equation~(\ref{eq:dxivec}) is linear in $\boldsymbol \xi(\tau)$, and otherwise it is very similar to Eq.~(\ref{eq:dmvec}). 
A known solution to Eq.~(\ref{eq:dxivec}) is the instanton solution ${\boldsymbol \xi}(\tau) = {\bf m}(\tau)$. Denote $\boldsymbol\xiplus(\tau)$ as the solution to Eq.~(\ref{eq:dxivec}) with the initial conditions 
\begin{align}\label{eq:initial_zeta}
 \xiplus_x(0) = 0\,,\quad i\xiplus_y(0) = \frac{1}{\sqrt 2}\,,\quad \xiplus_z(0) = \frac{1}{\sqrt 2}\,.
\end{align}
The corresponding state to $\xiplus_x(0)$ defined in Eq.~(\ref{eq:Xi}) is 
\begin{align}
  \ket{\bm\xiplus(0)}  = \frac{1}{\sqrt 2}\big(\ket{\Phi^+}+\ket{\Psi^+} \big)=  \ket{+}\otimes\ket{+}\,. 
\end{align}
Thus, the quantity in Eq.~(\ref{eq:trace_OB}) can be expressed as
\begin{align}\label{eq:trace_bilinear_form}
 \bra{+}K^{\beta}\ket{+}^2 = \bra{\bm\xiplus(0)} K^{\beta}\otimes K^{\beta} \ket{\bm\xiplus(0)} = \braket{\bm\xiplus(0)}{\bm\xiplus(\beta)}\,,
\end{align}
where $K^{\beta}$ is a shorthand for $K^{\beta, 0}$. To solve $\bm\xiplus(\beta)$, we define the real-valued symmetric bilinear form
\begin{align}
 \sB\big(\bm \xi(\tau), \bm\eta(\tau)\big) = \xi_x(\tau)\eta_x(\tau) +\xi_y(\tau)\eta_y(\tau) +\xi_z(\tau)\eta_z(\tau) \,,
\end{align}
where $\bm \xi(\tau) $ and $\bm\eta (\tau)$ are any two solutions to Eq.~(\ref{eq:dxivec}). Because the bilinear form is a constant of motion, we have the identities from the conditions~\eqref{eq:initial_instaton} and \eqref{eq:initial_zeta},
\begin{align}
 \sB\big(\boldsymbol \xiplus(\tau),\, \boldsymbol \xiplus(\tau)\big) = \sB\big(\boldsymbol \xiplus(\tau),\, {\bf m}(\tau)\big) = 0\;,\label{eq:pseudo_product}
\end{align}
for any $\tau\in[0,\beta]$. Consequently, we have
\begin{align}
 -\big(\xiplus_y^2+\xiplus_z^2\big) \mm_x^2 = \xiplus_y^2 \mm_y^2 +2 \xiplus_y \xiplus_z \mm_y \mm_z + \xiplus_z^2 \mm_z^2\;,
\end{align}
%\zj{The above equation has been modified.}
Introducing $\ratio = i\xiplus_y/ \xiplus_z$, we have
\begin{align}\label{eq:ratio_equation}
 (\ratio^2-1) \mm_x^2 = -\ratio^2 \mm_y^2 -2 i\ratio\, \mm_y \mm_z + \mm_z^2\,.
\end{align}
The solution to the above quadratic equation is
\begin{align}
 \ratio &= \frac{\mm_x-i\mm_y \mm_z}{1-\mm_z^2}\;,\label{eq:ratio}
\end{align}
where we use Eq.~(\ref{eq:int2}) to simplify things and the conditions~\eqref{eq:initial_zeta} [which implies $\ratio(0) = 1$] to rule out the other solution of Eq.~(\ref{eq:ratio_equation}).  Putting the condition~\eqref{eq:end_instanton} into Eq.~\eqref{eq:ratio}, we have 
\begin{align}\label{eq:ratio_end}
  \ratio(\beta) = \frac{1+\mm_z(\beta)^2}{1-\mm_z(\beta)^2}\,.
\end{align}
Combining Eqs.~\eqref{eq:pseudo_product} and \eqref{eq:ratio_end}, we have
\begin{gather}
  \xiplus_x(\beta) = -\sqrt 2\, \kappa\ssp \mm_z(\beta)\,,\\
  i\xiplus_y(\beta)= \kappa \big[1+\mm_z(\beta)^2\big]\big/\sqrt 2\,,\\
  \xiplus_z(\beta)= \kappa \big[1-\mm_z(\beta)^2\big]\big/\sqrt 2\,,\label{eq:xiplus_beta_z}
\end{gather}
where $\kappa$ is to be determined. Putting the definition of $\ratio$ into Eq.~(\ref{eq:dxivec}), we have
\begin{align}
 \dt \xiplus_z(\tau) = 2 \Gamma \ratio(\tau) \xiplus_z(\tau)\;,
\end{align}
%\zj{The sign of the above equation has been changed.}
which can be solved exactly,
\begin{align}
 \xiplus_z(\tau) = \xiplus_z(0)\, e^{2 \Gamma\! \int_0^\tau \ratio(\tauprim)\, \dif \tauprim}\;.\label{eq:xi_z}
\end{align}
Combining the above expression and Eq.~\eqref{eq:xiplus_beta_z}, we have
\begin{align}
 \ln \kappa = 2\Gamma\! \int_0^\beta \ratio(\tau)\, \dif \tau - \ln \big[1-\mm_z(\beta)^2\big]\,.
\end{align}
Putting Eq.~(\ref{eq:ratio}) into the above integral, we have
\begin{align}
\begin{split}
 \Gamma\int_0^\beta \ratio(\tau)\, \dif \tau
 &= \int_0^\beta \frac{\normb{\erg+ g(\mm_z)}-\mm_z \dt \mm_z/2}{1-\mm_z^2}\, \dif \tau\\
 &= \int_0^\beta \frac{ \normb{\erg+ g(\mm_z)}}{1-\mm_z^2}\, \dif \tau+\frac{1}{4} \ln \big(1-\mm_z^2\big)\Big\vert_0^\beta \,,
 \end{split}
\end{align}
%\zj{The first term has been changed.}
where the relations $\mm_x = \normb{\erg+ g(\mm_z)}/\Gamma$ and $i\mm_y = \dt \mm_z/2 \Gamma$ are used. The quantity $\kappa$ can thus be determined explicitly,
\begin{align}
 \ln \kappa = 2\mathcal I -\frac{1}{2}\Big( \ln \big[1-\mm_z(0)^2\big] + \ln \big[1-\mm_z(\beta)^2\big]\Big)\,,
\end{align}
where the integral $\mathcal I$ is defined as
\begin{align}\label{eq:sI}
 \mathcal I= \int_0^\beta \frac{ \norm{\erg+ g(\mm)}}{1-\mm^2}\, \dif \tau\,.
\end{align}
Noticing that $\kappa = \braket{\bm\xiplus(0)}{\bm\xiplus(\beta)}$ and using  Eqs.~\eqref{eq:trace_OB} and \eqref{eq:trace_bilinear_form}, we have
\begin{align}
 \widetilde\trprop = \tr( \omega K^{\beta}) &= 2\sqrt \kappa\,.
\end{align}
The free energy Eq.~(\ref{eq:F_OB}) can thus be evaluated,
\begin{align}
\begin{split}\label{eq:QMC_free_erg}
\beta F_\OB &=\int_{0}^{\beta}\big[m_z g'(m_z)-g(m_z)\big] \dif\tau
- \ln 2\\
&\quad - \mathcal I +\frac{1}{4}\Big( \ln \big[1-\mm_z(0)^2\big] + \ln \big[1-\mm_z(\beta)^2\big]\Big)\,.
\end{split}
\end{align}

\section{Wentzel-Kramers-Brillouin (WKB) approach}
\label{sec:wkb}

The partition function of the QMC Hamiltonian~\eqref{eq:partition_func} with OBC takes the form in the limit of large number of replicas
\begin{align}\label{eq:ob_partition_f_b}
 Z_\OB = \tr (\Omega\, e^{-\beta \hat H} )\,,
\end{align}
where $\hat H$ is defined in Eq.~\eqref{eq:H-sigma}, and $\Omega$ is the matrix of ones; i.e., $\Omega_{jk} = 1$ for all $j,k=1,2,\ldots, 2^N$. The rank of the matrix $\Omega$  is one, an it can be expressed as
\begin{align}
 \Omega = \proj{s}\,,
\end{align}
where $\ket{s} = \big(\ket{0}+\ket{1}\big)^{\otimes N}$ is in the symmetric subspace. The partition function~\eqref{eq:ob_partition_f_b} can also be written as
\begin{align}\label{eq:ob_partition_f_c}
 Z_\OB = \bra{s}  e^{-\beta \hat H} \ket{s}\,.
\end{align}
Letting $\ket{\mm}\equiv \ket{N/2,\, \mm N/2}$ denote the normalized state in the symmetric subspace of $N$ spin-1/2 particles with total magnetization $\mm N/2$, we have
\begin{align}\label{eq:wkb_partition_f}
 Z_\OB &=  \sum_{\mm_1,\mm_2}\braket{s}{\mm_2} \bra{\mm_2}e^{-\beta \hat H} \ket{\mm_1}\braket{\mm_1}{s}\,.
\end{align}
One distinction of QMC with OBC as opposed to PBC for the Hamiltonian~\eqref{eq:H-sigma} is that the discussion can be carried out in the symmetric subspace even at finite temperatures. The inner products $\braket{\mm_1}{s}$ and $\braket{s}{\mm_2}$ can be interpreted as additional \lq\lq entropic factors'' to the partition function. Indeed, we have
\begin{align}\label{eq:inner_Q}
\begin{split}
 \frac{\ln \braket{\mm}{s}}{N} &=  \frac{1}{N} \ln \sqrt{\frac{N!}{N_+!\, N_-!}}
% &\simeq \frac{1}{2N} \Big(N\ln N - N_+\ln N_+-N_-\ln N_-\Big)\\
\simeq \half\, \entropy(\mm)\,,
\end{split}
\end{align}
where $N_+ = N(1+\mm)/2$ and $N_- = N(1-\mm)/2$, and $\entropy(\mm)$ is the binary entropy defined in Eq.~(\ref{eq:entropy}). We will also need the derivative of the entropic factor,
\begin{align}\label{eq:deriv_Q}
 Q'(\mm) \equiv \frac{\dif Q(\mm)}{\dif \mm} = \frac{1}{2}\ln \frac{1-\mm}{1+\mm}\,.
\end{align}
The WKB \lq\lq free energy'' for OBC thus takes the form
\begin{align}
 \beta \mathcal{F}_\OB &=  \beta\erg+\frac{\action}{2}-\frac{1}{N}\Big(\ln\braket{\mm_1}{s} +\ln\braket{s}{\mm_2}\Big)\label{eq:free_erg_wkb_a} \\
 &\simeq \beta\erg+\frac{1}{2}\big(\action-Q(\mm_1)-Q(\mm_2)\big)\,,\label{eq:free_erg_wkb_b}
\end{align}
where $\erg$ is the energy. The WKB action $\action$ is given by the integral
\begin{align}
 \action &= \int_{\mm_1}^{\mm_2}  p\, \dif \mm\\
 &= \mm_2 p_2-\mm_1 p_1-\int_{\mm_1}^{\mm_2}  \mm p'\,\dif \mm\,,\label{eq:action_integral}
 \end{align}
where $p$ is the conjugate momentum of the magnetization $\mm$. The first two terms in Eq.~\eqref{eq:free_erg_wkb_a} are due to the propagator $\bra{\mm_2}e^{-\beta \hat H} \ket{\mm_1}$.
 
% We now show that the conditions~\eqref{eq:initial_instaton} and \eqref{eq:end_instanton} in Sec.~\ref{sec:open_boundary} can be derived by minimizing the WKB free energy $\mathcal{F}_\OB$. 
% The saddle point of the WKB free energy~\eqref{eq:free_erg_wkb_b} is the instanton solution given by the Hamiltonian
% \begin{align}
%  H_\WKB(p,\mm) = -\Gamma\sqrt{1-\mm^2}\cos p -g(\mm)\,.
% \end{align}
% A small change in energy takes the form
% \begin{align}
% \begin{split}
%  \delta H_\WKB(p,\mm)  &= \frac{\partial H_\WKB}{\partial p}\,\delta p + \frac{\partial H_\WKB}{\partial \mm}\, \delta \mm\\
%  &= -\half\big(\dt \mm\,\delta p -\dt p\, \delta \mm\big)\,.
% \end{split}
% \end{align}
%\zj{The factor -1/2 is due to different definitions.}
A small change in energy of the path takes the form
\begin{align}
  \delta \erg &= -\frac{1}{2\beta}\int_0^\beta \big(\dt \mm\,\delta p -\dt p\, \delta \mm\big)\,\dif \tau\\ 
  % &= -\frac{1}{2\beta} \,\delta\! \int_0^\beta \dt \mm p \,\dif \tau + \frac{p\ssp \delta \mm}{2\beta}\,\Big\rvert_0^\beta\\
  &= -\frac{\delta \action}{2\beta} +\frac{1}{2\beta}\,(p_2\delta \mm_2-p_1\delta\mm_1)\,.\label{eq:delta_erg}
\end{align}
% Fixing $\mm_1$ and $\mm_2$ ($\delta \mm_1 = \delta\mm_2 = 0$), we have
% \begin{align}
%  \frac{\partial \mathcal{F}_\OB}{\partial e} = 
%  1+\frac{1}{2\beta}\ssp\frac{\partial \action}{\partial e} = 0\,,
% \end{align}
% which can be used to determine the energy $\erg$.
Using Eqs.~\eqref{eq:free_erg_wkb_b} and \eqref{eq:delta_erg}, we have
\begin{align}
\begin{split}
  \delta\mathcal{F}_\OB 
 &\simeq \delta \erg+\frac{1}{2 \beta}\big(\delta\action-\delta Q(\mm_1)-\delta Q(\mm_2)\big)\\
 &= \frac{1}{2\beta}\,\Big( [p_2-Q'(\mm_2)]\delta \mm_2-[p_1+Q'(\mm_1)]\delta\mm_1\Big)\,.
  \end{split}
\end{align}
% Fixing $\erg$, we have 
% \begin{align}\label{eq:F_boundary}
%  \frac{\partial \mathcal{F}_\OB}{\partial \mm_1} = \frac{\partial \mathcal{F}_\OB}{\partial \mm_2} = 0\,.
% \end{align}
% Putting Eqs.~\eqref{eq:free_erg_wkb_b} and \eqref{eq:delta_erg} into Eq.~\eqref{eq:F_boundary}, we have
The boundary conditions of any saddle point of $\mathcal{F}_\OB$ can then be derived,
\begin{align}\label{eq:end_points}
 p_1 + Q'(\mm_1) = p_2 -Q'(\mm_2) = 0\,. 
\end{align}
We now show that these conditions are identical to Eqs.~\eqref{eq:initial_instaton} and \eqref{eq:end_instanton}. The WKB instanton solution is given by [same as Eq.~(\ref{eq:mx})]
\begin{align}\label{eq:momentum}
 \norm{p} = \cosh^{-1}  \frac{\normb{\erg+g(\mm)}}{\Gamma \sqrt{1-\mm^2}}\;.
\end{align}
Using conditions~\eqref{eq:initial_instaton} and \eqref{eq:end_instanton}, we have $\normb{\erg+g(\mm)}/\Gamma = \mm_x = 1$ for the end points. Putting this into Eq.~\eqref{eq:momentum}, we recover the results in Eq.~\eqref{eq:end_points},
\begin{align}\label{eq:boundary_p}
 \norm {p}= \cosh^{-1}\frac{1}{\sqrt{1-\mm^2}} =  \half \ln \frac{1 +\norm{\mm}}{1-\norm{\mm}} = \big\lvert Q'(\mm) \big\rvert
 \,.
\end{align} 

For the instanton solution, the integral in Eq.~\eqref{eq:action_integral} takes the form
\begin{align}
\int_{\mm_1}^{\mm_2}&  \mm p'\,\dif \mm\nonumber\\ 
 &= \int_{\mm_1}^{\mm_2}  \frac{-\mm g'(\mm)+ \frac{\mm^2}{1-\mm^2}\normb{\erg+g(\mm)}}{\sqrt{[\erg+g(\mm)]^2 -\Gamma^2 \big(1-\mm^2\big)}} \,\norm{\dif \mm} \\ 
 &= 2\int_{0}^{\beta}\Big(\mathord -\mm g'(\mm) + \frac{\mm^2}{1-\mm^2}\normb{\erg+g(\mm)}\Big)\,\dif \tau\\
 &= 2\big(\beta \erg+\mathcal I\big)-2\int_{0}^{\beta} \big[\mm g'(\mm)-g(\mm)\big]\,\dif \tau \,,\label{eq:wkb_action}
\end{align}
where we use the instanton solution given in Eq.~\eqref{eq:instanton_eq}, and $\mathcal I$ is defined in Eq.~\eqref{eq:sI}. For $\mm_1<0$ and $p_1<0$, we have 
\begin{align}
 \mm_1 p_1 &= \frac{\mm_1}{2} \ln \frac{1+\mm_1}{2}- \frac{\mm_1}{2}\ln \frac{1-\mm_1}{2}\\
 &= -Q(\mm_1) -\half \ln(1-\mm_1^2)+\ln 2 \,.
\end{align}
For $\mm_2>0$ and $p_2<0$, we have 
\begin{align}
 \mm_2 p_2 &= \frac{\mm_2}{2} \ln \frac{1-\mm_2}{2}- \frac{\mm_2}{2}\ln \frac{1+\mm_2}{2}\\
 &=Q(\mm_2) +\half \ln(1-\mm_2^2)-\ln 2 \,.
\end{align}
Putting the above results together, we have
\begin{align}\label{wkb_action_b}
\begin{split}
 \action &= \mm_2 p_2-\mm_1 p_1-\int_{\mm_1}^{\mm_2}  \mm p'\,\dif \mm\\
 &=Q(\mm_1)+ Q(\mm_2) + \frac{1}{2} \big[\ln(1-\mm_1^2)+\ln(1-\mm_2^2)\big] \\
 &\quad -2\big(\ln 2 +\beta \erg+\mathcal I\,\big)+2\int_{0}^{\beta} \big[\mm g'(\mm)-g(\mm)\big]\,\dif \tau\,.
 \end{split}
\end{align}
Putting Eq.~\eqref{wkb_action_b} into \eqref{eq:free_erg_wkb_b}, we have the effective WKB \lq\lq free energy''
\begin{align}\label{eq:free_erg_wkb}
\begin{split}
 \beta \mathcal{F}_\OB 
%  &\equiv  \beta\erg+\frac{\action}{2}-\frac{1}{N}\Big(\ln\braket{\mm_1}{s} +\ln\braket{s}{\mm_2}\Big) \\
%  &= \beta\erg+\frac{1}{2}\big(\action-Q(\mm_1)-Q(\mm_2)\big)\\
 &= \int_{0}^{\beta}\big[\mm g'(\mm)-g(\mm)\big] \dif\tau
- \ln 2\\
&\quad - \mathcal I +\frac{1}{4}\Big( \ln \big[1-\mm_1^2\big] + \ln \big[1-\mm_2^2\big]\Big)\;,
  \end{split}
\end{align}
which is identical to the QMC free energy $F_\OB$ in Eq.~(\ref{eq:QMC_free_erg}).

Another surprising fact about the QMC with OBC is that the minimum of the free-energy functional is not achieved by a static solution independent of $\tau$. This can be understood by the definition of $\mathcal{F}_\OB$ in Eq.~\eqref{eq:free_erg_wkb}; while $\action$ and $\erg$ can be minimized simultaneously by an optimal static solution, the boundary terms are not minimized by that solution. As a consequence, the free-energy functional is minimized by a nontrivial solution caused by the entropic factors at the boundaries. The end points of this instanton satisfy $\mm_1=\mm_2$ and $p_1=-p_2$. To get rid off this effect, one can add boundary terms (pinning fields) to cancel the extra entropic factors. 

\section{Spin-coherent-state quantum Monte Carlo}

The instantonic analysis presented here allows the study of spin tunneling for a class of Hamiltonians, symmetric with respect to permutations of individual spin-1/2 particles (qubits).
It can potentially be generalized to the case without permutation symmetry using path integrals over spin-coherent states.
The action along the instanton trajectory in imaginary time is
\begin{align}
 \action ({\bf n})=\frac{i\hbar}{2}\sum_{i=1}^{N} \chi({\bf n}_j)+\int_{0}^{\beta} \dif\tau\,  H[{\bf n}_1(\tau),\ldots,{\bf n}_N(\tau)],\label{eq:Action}
\end{align}
where ${\bf n}_j(\tau)$ is the Bloch unit vector of the $j$th spin and the Berry phases of individual spins are defined as
 \begin{equation}
   \chi({\bf n}_j)=\int_{0}^{\beta}\dif\tau \big[1-\cos\theta_j(\tau)\big]\dot \phi_j(\tau)\;,
   \end{equation} 
where $\theta_j(\tau)$ and $\phi_j(\tau)$ are the polar angle and azimuthal angle of the Bloch unit vector ${\bf n}_j(\tau)$, respectively. The function $H[{\bf n}_1(\tau),\ldots,{\bf n}_N(\tau)]$ in (\ref{eq:Action}) is obtained from the system Hamiltonian $H$ by replacing the Pauli matrices of the $j$th spin by the 
corresponding components of the Bloch unit vector ${\bf n}_j(\tau)$. 

Within the incoherent tunneling framework, the instanton trajectory with open-boundary condition connects the spin coherent state $\{{\bf n}_j(0)\}_{j=1}^{N}$ corresponding to the maximum of the wavefunction near the local minima of the energy landscape $H[{\bf n}_1(\tau),\ldots,{\bf n}_N(\tau)]$ to the coherent state $\{{\bf n}_j(\beta)\}_{j=1}^{N}$ corresponding to the remote tail of the wavefunction on the other side of the barrier. Such a tunneling transition corresponds to correlated motions of individual spins in imaginary time satisfying $\delta \action/\delta {\bf n}_j(\tau)=0$ (to minimize the action). Therefore, the Bloch unit vector evolves according to the instanton equation,
\begin{equation}
\frac{\hbar}{2}\frac{d{\bf n}_j(\tau)}{d\tau}={\bf n}_j(\tau)\times\frac{\partial H}{\partial {\bf n}_j(\tau)}.\label{eq:dndt}
\end{equation}
We note that the first (Berry phase) term in (\ref{eq:Action}) contains additional factor $i$ compared to the second term. Therefore, the $y$ component of ${\bf n}_j(\tau)$ is complex for the instanton trajectory [cf. Eqs.~(\ref{eq:mvec})--(\ref{eq:dt_mz}) above]. The Bloch unit vector can be written in the form 
 \begin{equation}
 {\bf n}_j =(\sin\theta_j \cosh\varphi_j,  -i \sin\theta_j \sinh\varphi_j, \cos\theta_j)\;,\label{eq:n}
\end{equation}
corresponding to a purely imaginary azimuthal angle $\phi_j(\tau)=-i \varphi_j(\tau)$. This substitution makes the Berry phase terms in Eq.~(\ref{eq:Action}) real along the instanton path. The values of $H[{\bf n}_1(\tau),\ldots,{\bf n}_N(\tau)]$ are also real due to the fact that the Hamiltonian $H$ is Hermitian. Therefore, despite the presence of the imaginary Berry phase in Eq.~(\ref{eq:Action}), the instanton trajectory equations (\ref{eq:dndt}) involve only real quantities after the substitution of Eq.~(\ref{eq:n}). A similar situation also happens in  the spin tunneling studied above via  Eqs.~(\ref{eq:mvec})--(\ref{eq:dt_mz}).
Finally, the tunneling matrix element $\Delta_{\rm tunn}$ is determined with logarithmic equivalence as 
\begin{equation}
 \Delta_{\rm tunn} = \B \exp[-\action({\bf n}^*)]\;,\label{eq:Wgen}
\end{equation}
where ${\bf n}^*$ is an instanton trajectory and the prefactor $\B$ can be obtained in terms of the functional determinant of the kernel $\delta^2 \action({\bf n})$ at ${\bf n}^*$.
 
The variational method outlined above can be used to study the tunneling matrix elements in transverse-field spin glasses between the computational basis states separated by large Hamming distances. These matrix elements are usually exponentially small in the number of spins $N$, and alternative methods involving exact diagonalization are not feasible for $n\gtrsim 30$.  

Furthermore, one can think of the above method as a basis for an alternative QMC approach, where one directly samples the paths $\{\theta_j(\tau),\varphi_j(\tau)\}_{j=1}^{N}$ in imaginary time according to the probability functional $\propto \exp[-\action({\bf n})]$. The explicit form of the prefactor can be obtained using the results in Ref.~\cite{garg2003spin}. The trajectories of two variables for each spin need to be simulated in such Monte Carlo methods which avoids topological obstructions such as those considered in Ref.~\cite{andriyash_can_2017}.

\section{Discussion and Summary}

In QMC, quasi-equilibrium distributions of paths are determined by classical free-energy functionals. Transitions from a local minimum to the global minimum are described by Kramers escape events, where the system reaches a single \lq\lq transition state" (a saddle point) before making an escape from the metastable state. We found the transition state in open-boundary QMC simulations analytically for a fully connected spin model with bit-symmetric cost functions. The transition state corresponds to an instanton, governed by the same differential equation as in the PBC case~\cite{jiang_scaling_2017}. We derive the instanton equation using two different approaches. One is the mean-field approach, where the free-energy functional is expressed using a two-dimensional propagator. This propagator corresponds to a particle evolving in imaginary time under the action of a time-dependent magnetic field. The instanton equation can be derived by observing two constants of motion. Another approach is to map the free energy of the open-boundary QMC to an imaginary-time quantum propagator $\bra{s}  e^{-\beta \hat H} \ket{s}$, where $\ket{s} = \big(\ket{0}+\ket{1}\big)^{\otimes N}$; this contrasts to the PBC case, where the free energy is related to $\tr  e^{-\beta \hat H} $. The instanton solution that dominates the escape event can be derived using the WKB approach in the large-spin picture, which coincides with our mean-field results. The instanton in periodic-boundary QMC is determined by the instanton equation and the period (inverse temperature); however, one also has to optimize the positions and momenta at both ends of an instanton in open-boundary QMC. We found that the initial and final momenta of instantons in open-boundary QMC are nonzero; this is because the extra entropic factors introduced at the open boundaries. To cancel this side effect, one can add extra potential terms at the end replicas to suppress configurations with higher entropy. At finite temperatures, the periodic-boundary QMC escape event corresponds to a quantum tunneling path in a subspace with total angular momentum less than the maximum value~\cite{jiang_scaling_2017}. With OBCs, however, this result no longer holds; we found that the QMC escape event always corresponds to tunneling events in the symmetric subspace (maximum total angular momentum). We also outlined the generalization of the instantonic tunneling method to systems without permutation symmetry using spin-coherent-state path integrals.

\section{Acknowledgments}

Z.J. would like to acknowledge enlightening and useful discussions with Salvatore Mandr\`{a} and Andre Petukhov. This work is supported by the NASA Advanced Exploration Systems program and NASA Ames Research Center. The research is based in part upon work supported by the Office of the Director of National Intelligence (ODNI), Intelligence Advanced Research Projects Activity (IARPA), via Interagency Umbrella Agreement No. IA1-1198. The views and conclusions contained herein are those of the authors and should not be interpreted as necessarily representing the official policies or endorsements, either expressed or implied, of the ODNI, IARPA, or the U.S. Government. The U.S. Government is authorized to reproduce and distribute reprints for Governmental purposes notwithstanding any copyright annotation thereon.

% \bibliographystyle{apsrev4-1_with_title}
% \bibliography{bibQA}


\appendix

\section{Meanfield description of path-integral quantum Monte Carlo}
\label{sec:PIQMC}

The partition function of $H_\QMC$ defined in Eq.~(\ref{eq:partition_func}) can be calculated with a mean-field approach. First, we consider $N$ noninteracting spins under the mean-field Hamiltonian
\begin{align}\label{eq:hamiltonian_local_field}
 H_\MF = -\frac{N}{R}\sum_{\tau=0}^{\nrepl-1} \mfield(\tau)  \mm(\tau)-J\sum_{j=1}^{N}\sum_{\tau=0}^{\nrepl-1}\sigma_j(\tau)\sigma_j(\tau+1)\,,
\end{align}
where $\mfield(\tau)$ is an \lq\lq average'' local magnetic field (mean-field) on the $\tau$th replica slice and $\mm(\tau)$ is defined in Eq.~(\ref{eq:m_tau}). The partition functional corresponds to $H_\MF$ is
\begin{align}\label{eq:mf_partition_f}
 \mathcal Z_\MF(\bm \lambda) = [\trprop(\bm \lambda)]^N\;,
\end{align}
where $\bm \lambda = \big(\lambda(0),\ldots, \lambda(\nrepl-1)\big)$. The function $\trprop(\bm \lambda)$ is defined as
\begin{align}
   \trprop(\bm \lambda)&=\sum_{\sigma_0,\ldots,\sigma_{\nrepl-1}} e^{\beta \sum\nolimits_{\tau=0}^{\nrepl-1}\big( \mfield(\tau)\ssp \sigma(\tau)/\nrepl + J\sigma(\tau)\sigma(\tau+1)\big)}\nonumber\\
    &= \tr \Big(L[\lambda(\nrepl-1)]\cdots  L[\lambda(\tau)] \cdots L[\lambda(0)]\Big)\,,\label{eq:trprop}
\end{align}
where the transition matrix takes the form
\begin{align}\label{eq:G_matrices}
 L(\lambda)=
 \begin{pmatrix}
  e^{\beta J} & e^{-\beta J}\\
  e^{-\beta J} & e^{\beta J}
 \end{pmatrix}
 \begin{pmatrix}
  e^{\beta\mfield/\nrepl} & 0\\
  0 & e^{-\beta \mfield/\nrepl}
 \end{pmatrix}\;.
\end{align}
The trace in Eq.~(\ref{eq:trprop}) is a consequence of the periodic-boundary condition $\sigma_j(0)=\sigma_j(\nrepl-1)$ typically used in QMC. By defining the propagator
\begin{align}\label{eq:K}
K^{\tau_2,\tau_1} = L [\lambda(\tau_2)] \cdots L[\lambda(\tau_1)]\,,\quad \tau_2> \tau_1\,,
\end{align}
we have
\begin{align}\label{eq:trprop2}
\trprop(\bm \lambda)= \tr K^{\nrepl-1,\, 0}(\bm \lambda)\,.
\end{align}
The expectation value of $\mm(\tau)$ can be calculated from the partition function
\begin{align}\label{eq:m_B}
\begin{split}
 \overline \mm(\tau)  &= \frac{\nrepl}{\beta N}\frac{\partial \ln \mathcal \mathcal Z_\MF(\bm \lambda)}{\partial \mfield(\tau)}\\
  &=\frac{\nrepl}{\beta }\frac{\partial \ln \trprop(\bm \lambda)}{\partial \mfield(\tau)} = \frac{\tr \big[K^{\nrepl-1,\, \tau} \sigma_z\,  K^{\tau-1,\, 0}  \big]}{\trprop(\bm \lambda)}\,,
  \end{split}
\end{align}
where $\sigma_z$ is the Pauli-$z$ matrix. The expectation values of the magnetization $ \overline \mm(0),\ldots, \overline \mm(\nrepl-1)$ are functions of the parameters $\mfield(0),\ldots,\mfield(\nrepl-1)$, and vice versa. Since the fluctuation of $\mm(\tau)$ vanishes in the large $N$ limit, we neglect the difference between  $\overline{\mm}(\tau)$ and $\mm(\tau)$, and Eq.~(\ref{eq:m_B}) becomes
\begin{align}\label{eq:m_lambda}
 \mm(\tau) = \frac{\nrepl}{\beta }\frac{\partial \ln \trprop(\bm \lambda)}{\partial \mfield(\tau)} \,.
\end{align}
The free-energy functional corresponding to the Hamiltonian~(\ref{eq:hamiltonian_local_field}) is
\begin{align}
 \free_\MF = -\frac{1}{\beta N} \ln \mathcal Z_\MF (\bm \lambda )= -\frac{1}{\beta} \ln \trprop(\bm \lambda )\,.
\end{align}
Consequently, the free-energy functional of the QMC Hamiltonian~(\ref{eq:hamiltonian}) reads (see also Ref.~\cite{jorg2010energy}),
\begin{align}\label{eq:free_erg}
\begin{split}
 \free  &= \frac{1}{\nrepl}\sum_{\tau=0}^{\nrepl-1} \Big(\mfield(\tau) \mm(\tau) -\g[\mm(\tau)] \Big)-\frac{1}{\beta} \ln \trprop(\bm \lambda )\,,
\end{split}
\end{align}
where the term $\mfield(\tau) \mm(\tau)$ is introduced to cancel the corresponding term in Eq.~(\ref{eq:hamiltonian_local_field}). When the QMC free-energy functional $\free$ is minimized, we have
\begin{align}
 \nrepl\,\frac{\partial \free}{\partial \mm(\tau)} &=\mfield(\tau)  -\g'[\mm(\tau)]\nonumber\\
 &\quad + \sum_{\tauprim=0}^{\nrepl-1} \bigg(\mm(\tauprim)\frac{\partial \mfield(\tauprim)}{\partial \mm(\tau)}- \frac{\nrepl}{\beta }\frac{\partial \mfield(\tauprim)}{\partial \mm(\tau)} \frac{\partial \ln \trprop( \bm \lambda ) }{\partial \mfield(\tauprim)} \bigg)\nonumber\\[2pt]
 &= \mfield(\tau)  -\g'[\mm(\tau)]=0\,,\label{eq:saddle_point}
\end{align}
where Eq.~(\ref{eq:m_lambda}) is used to show that the sum over $\tauprim$ equals to zero. Equations~(\ref{eq:m_lambda}) and (\ref{eq:saddle_point}) determine the saddle point of the free-energy functional~(\ref{eq:free_erg}). 

To calculate the exponential scaling of the QMC escape rate using the instanton method, we will need to go the continuous limit $\nrepl\rightarrow \infty$. In this limit, the propagator defined in Eq.~(\ref{eq:K}) corresponds to a spin-1/2 particle evolving in imaginary time under the action of the time-dependent magnetic field $\mathbf B(\tau)$,
\begin{gather}
% \widetilde\trprop(\bm \lambda) =\tr \big(\omega K^{\beta,0}\big),\label{eq:Lambda}\\
K^{\tau_2,\tau_1} ={\rm T_{+}}e^{-\int_{\tau_1}^{\tau_2} d\tau  H_0(\tau)}\,,\label{eq:propagator_app}\\
H_0(\tau) =-\mathbf{B}(\tau)\cdot \bm{\sigma},\quad \mathbf B(\tau)=\big(\Gamma,0,\lambda(\tau)\big)\label{eq:H0_app}\;,
\end{gather}
where $\tau\in[0,\,\beta]$ denotes the imaginary time, ${\rm T_{+}}$ is the time-ordering operator, and $\bm \sigma=(\sigma_x,\sigma_y,\sigma_z)$ is the vector of Pauli matrices. We will replace the notations $K^{\tau_2,\tau_1}(\bm\lambda)$ and $\trprop(\bm\lambda)$ with $K^{\tau_2,\tau_1}(\lambda)$ and $\trprop(\lambda)$, because $\lambda(\tau)$ is a function in the continuous limit.

Similar to Eqs.~(\ref{eq:mf_partition_f}) and (\ref{eq:trprop2}), the meanfield partition functional with open-boundary condition takes the form in the continuous limit
\begin{align}\label{eq:ob_partition_f_app}
   \mathcal Z_\MF^\OB = [\widetilde\trprop(\lambda)]^N\,,\quad \widetilde\trprop(\lambda) =\tr \big(\omega K^{\beta,\, 0}(\lambda) \,\big)\,,
\end{align}
where the matrix
\begin{align}
 \omega=1+\sigma_x=\begin{pmatrix}1&1\\1&1\end{pmatrix}\,.
\end{align}
The matrix $\omega$ makes it possible to sum over configurations prohibited by the periodic-boundary condition, which is unique to open-boundary QMC. Similar to Eq.~(\ref{eq:free_erg}), the free-energy functional of open-boundary QMC takes the form,
\begin{align}\label{eq:F_OB_app}
\free_\OB=\frac{1}{\beta} \int_{0}^{\beta}\Big(\mm(\tau) \lambda(\tau)-g[\mm(\tau)]\Big) d\tau
- \frac 1 \beta \ln \widetilde\trprop(\lambda)\,.
\end{align}
The saddle-point conditions for open-boundary QMC are similar to those for periodic-boundary QMC; Eq.~(\ref{eq:saddle_point}) remains the same, and one just need to replace $\trprop$ with $\widetilde\trprop$ in Eq.~(\ref{eq:m_lambda}), 
\begin{equation}\label{eq:dm-var_app}
\lambda(\tau)= g^\prime[\mm(\tau)]\,,\quad \mm(\tau)=\frac{\delta \ln \widetilde\trprop(\lambda)}{\delta \lambda(\tau)}\,.
\end{equation}

% \begin{align}
% \free_\OB=\frac{1}{\beta} \int_{0}^{\beta}\Big(\mm(\tau) \lambda(\tau)-g[\mm(\tau)]\Big) d\tau
% - \frac 1 \beta \ln \widetilde\trprop(\lambda)\,.
% \end{align}

\end{document}